\documentclass[pra,preprint,aps,showpacs]{revtex4}
\large
\usepackage{epsfig}
\begin{document}
\draft
\input{epsf}
\title{Radiative friction on an excited atom moving in vacuum}
\author{Wei Guo\footnote{Electronic address: guow@queens.edu}}
\affiliation{Department of Natural Sciences, Queens University of 
             Charlotte, 1900 Selwyn Avenue, Charlotte, North Carolina 
             28274, USA} 
\date{\today}
\begin{abstract}
  It is known that, when an excited atom spontaneously emits one  
  photon, two effects are produced.  First, the atom's internal and 
  external states are entangled with the states of the emitted 
  photon.  Second, the atom receives a momentum transferred from the 
  photon.  In this work, the dynamics of such an atom in vacuum is 
  studied.  Through a specific calculation, it is demonstrated that 
  these effects cause the atom to experience, on average, a friction 
  force opposite to its initial velocity.  Properties of the force are 
  also discussed.
\end{abstract}
\maketitle
  Because of atom-vacuum interaction, an excited atom is able to emit 
  a photon spontaneously into any direction \cite{man95}, and 
  consequently has its internal and external states entangled with 
  photonic states \cite{rz92,guo10}.  Thus, even inside the vacuum, the
  atom's motion can never be a free motion.  Already pointed out was 
  that, in the presence of such photon-atom entanglement, the 
  atom's position-momentum uncertainty can be temporarily lower than 
  the lower bound specified by the Heisenberg uncertainty relation 
  \cite{guo10}.  It is certainly desirable to know what other 
  effects the atom-vacuum interaction can have on the atom's motion.
  
  Note, the frequency of the emitted photon depends on the direction 
  in which the photon is emitted according to the Doppler effect.  In 
  particular, when the photon is in the same direction as the atom's 
  velocity, it must have a frequency larger than the frequency it has 
  when emitted in the opposite direction.  On the other hand, the 
  photon emitted in any two directions perpendicular to the velocity 
  must have the same frequency.  Since a higher frequency means 
  that a larger momentum is carried away by the photon, and that a 
  larger momentum is transferred to the atom from the photon, the 
  excited atom, when moving, must receive, on average, a nonzero 
  momentum transferred from the photon, and should experience, in 
  addition to the photon-atom entanglement, a radiative force, on 
  average, opposite to its velocity.  The present paper is devoted to 
  the derivation of the radiative force $\vec{F}_{r}$.  Since the 
  radiative force is opposite to the atom's velocity, it is addressed 
  as radiative friction in the paper.  In the literature, radiative 
  forces on an atom are often discussed when the atom is exposed to 
  laser beams \cite{da85,man95}, and are seldom analyzed when the atom 
  is simply in the vacuum.  Unlike the laser beams, the vacuum has far 
  more electromagnetic modes to be considered.  It will become evident 
  in the following discussion that $\vec{F}_{r}$ also depends 
  critically on the photon-atom entanglement.   
   
  Denote the momentum operator, position operator, and mass of the 
  excited atom as $\vec{P}$, $\vec{R}$, and $m$ respectively.  For 
  simplicity, the atom is assumed to have two energy states: an excited
  state $\vert E \rangle$ with energy $\hbar \omega _{E}$ and the 
  ground state $\vert G \rangle$ with energy $\hbar \omega _{G}$.  The 
  difference between $\omega _{E}$ and $\omega _{G}$ is known as the 
  atomic transition frequency $\omega _{0}\equiv \omega _{E}- 
  \omega _{G}$.  The interaction between the vacuum modes and atom is 
  assumed, as usual, to be through the atom's electric dipole moment 
  $\vec{\mu}$ and is described by an operator $V_{I}$ defined as 
  \begin{equation} 
  \label{e1}
  V_{I}=\sum _{\alpha}\Big (\vec{\mu}_{GE}\cdot \vec{g} _{\alpha}e^{-i 
  \vec{k}_{\alpha}\cdot \vec{R}}a^{\dagger}_{\alpha}\vert G \rangle
  \langle E \vert +\vec{\mu}_{EG}\cdot \vec{g}^{\ast}_{\alpha}
  e^{i\vec{k}_{\alpha}\cdot \vec{R}} a_{\alpha} \vert E \rangle 
  \langle G \vert \Big ),
  \end{equation}
  where $\vec{\mu}_{GE}$ is the matrix element of $\vec{\mu}$ between 
  $\vert G \rangle$ and $\vert E \rangle$, and $\vec{\mu}_{EG}$ the 
  complex conjugate of $\vec{\mu}_{GE}$.  The frequency and amplitude 
  (containing a polarization unit vector $\vec{\epsilon}_{\alpha}$) of 
  mode $\alpha$ are represented by $\omega _{\alpha}$ and 
  $\vec{g}_{\alpha}=i\sqrt{2\pi \hbar \omega ^{2}_{0}/(L^{3}
  \omega _{\alpha})} \vec{\epsilon}_{\alpha}$ respectively.  The 
  quantization volume is $L^{3}$.  Also used in $V_{I}$ are 
  $\vec{k}_{\alpha}$, the wave vector of mode $\alpha$ 
  ($\mid \vec{k}_{\alpha}\mid =\omega _{\alpha}/c$), and 
  $a^{\dagger}_{\alpha}$ ($a_{\alpha}$), the creation (annihilation) 
  operator for the same mode.  Throughout the paper, $c$ is the speed 
  of light in the vacuum, and $\ast$ denotes the complex conjugate of a
  quantity.  Still for simplicity, $V_{I}$ is treated 
  under the rotating-wave approximation \cite{R1}.  The Hamiltonian $H$
  of the atom-vacuum system is constructed by adding to $V_{I}$ the 
  unperturbed Hamiltonian $H_{0}=P^{2}/(2m)+\hbar 
  \omega _{E} \vert E \rangle \langle E \vert +\hbar \omega _{G} 
  \vert G \rangle \langle G \vert +\sum _{\alpha}\hbar \omega _{\alpha} 
  a^{\dagger}_{\alpha}a_{\alpha}$:
  \begin{equation}
  \label{e2}
  H=H_{0}+V_{I}.
  \end{equation}
  Since it is constant and unimportant to the atom-vacuum system's 
  evolution, the zero point energy of the vacuum is ignored in 
  $H$.  Also ignored is the R\"{o}ntgen interaction, because this 
  interaction \cite{wi94,bo02} is roughly of the order of $vc^{-1}
  V_{I}$ (where $v$ is the atom's speed), much weak compared with 
  $V_{I}$ in the present nonrelativistic analysis.  For example, in a 
  typical experiment on spontaneous emission \cite{ku97}, $v$ is 
  merely of the order of $10^{3}m/s$.

  The atom's initial external state $\vert \psi (0) \rangle =\int 
  d\vec{p}_{0}f(\vec{p}_{0}) \vert \vec{p}_{0}\rangle$ is 
  expressed in terms of the eigenstates $\vert \vec{p}_{0} \rangle$ of 
  the momentum operator.  To mimic the initial condition, assumed 
  without loss of generality, that the atom moves along the positive 
  direction $\hat{x}$ of the $x$-axis of a coordinate system stationary
  in the vacuum, the eigenvalues $\vec{p}_{0}$ 
  of these states are assumed to be all along $\hat{x}$.  Since the 
  atom is in the excited state $\vert E \rangle$, and no photons are 
  present $\vert 0 \rangle$, evolution of the atom-vacuum system must 
  start from such an initial state $\vert \phi (0) \rangle = \vert 
  \psi (0) \rangle \otimes \vert E \rangle \otimes \vert 0 
  \rangle$.  At time $t$, the state of the system $\vert \phi (t) 
  \rangle$ consequently becomes
  \begin{equation}
  \label{e3}
  \vert \phi (t) \rangle =e^{-\frac{iHt}{\hbar}}\vert \phi (0) 
  \rangle =-\frac{1}{2\pi i}\int ^{\infty}_{-\infty} 
  dq \frac{e^{-iqt/\hbar }}{q-H} \vert \phi (0) \rangle .
  \end{equation} 
  As, for example, in the discussion of the Ehrenfest theorem 
  \cite{sa94} and the atomic motion in the laser beams 
  \cite{da85,man95}, the radiative friction $\vec{F}_{r}$ is defined 
  to be proportional to the second-order time derivative of the 
  expectation value of the position operator $\vec{R}$ taken with 
  respect to state $\vert \phi (t) \rangle$:
  \begin{equation}
  \label{e4}
  \vec{F}_{r}=m\frac{d^{2}}{dt^{2}}\langle \phi (t) \vert \vec{R} \vert
  \phi (t) \rangle .
  \end{equation}
  
  Consider the $x$-component of the force $\vec{F}_{r}$.  From 
  Eq. (\ref{e3}), the time-dependent expectation value of $R_{x}$ (the 
  $x$-component of $\vec{R}$),
  \begin{equation}
  \label{e5}
  \langle R_{x} \rangle (t)=\langle \phi (t) \vert R_{x} \vert \phi (t) 
  \rangle= \langle \phi (0) \vert e^{\frac{iHt}{\hbar}} R_{x} 
  e^{-\frac{iHt}{\hbar}} \vert \phi (0) \rangle , 
  \end{equation} 
  and its first-order derivative, 
  \begin{eqnarray}
  \label{e6}
  \frac{d}{dt} \langle R_{x} \rangle (t)&=&\frac{i}{\hbar} \langle 
  \phi (0) \vert e^{\frac{iHt}{\hbar}}[H,R_{x}] 
  e^{-\frac{iHt}{\hbar}} \vert \phi (0) \rangle \nonumber\\
  &=& \frac{1}{m} \langle \phi (0) \vert e^{\frac{iHt}{\hbar}} 
  P_{x} e^{-\frac{iHt}{\hbar}} \vert \phi (0) \rangle ,
  \end{eqnarray}
  are first obtained.  In Eq. (\ref{e6}), $P_{x}$ is the $x$-component 
  of the momentum operator $\vec{P}$.  From the preceding equation, it 
  is then a straightforward matter to find the second-order time 
  derivative of $\langle R_{x} \rangle (t)$: 
  \begin{eqnarray}
  \label{e7}
  \frac{d^{2}}{dt^{2}} \langle R_{x} \rangle (t)&=& \frac{i}{m} 
  \langle \phi (t) \vert \sum _{\alpha} (\vec{\mu}_{GE}\cdot 
  \vec{g}_{\alpha})a^{\dagger}_{\alpha} \vert G \rangle 
  \langle E \vert k_{\alpha x} e^{-i\vec{k}_{\alpha}\cdot \vec{R}}
  \vert \phi (t) \rangle \nonumber\\
  & &- \frac{i}{m} \langle \phi (t) \vert \sum _{\alpha} 
  (\vec{\mu}_{EG}\cdot \vec{g}^{\ast}_{\alpha})a_{\alpha} \vert E 
  \rangle \langle G \vert k_{\alpha x} 
  e^{i\vec{k}_{\alpha}\cdot \vec{R}} \vert \phi (t) \rangle ,
  \end{eqnarray}
  where $k_{\alpha x}$, as the $x$-component of the mode vector 
  $\vec{k}_{\alpha}$, illustrates the momentum exchange between the 
  photon and atom when the photon is created or annihilated. 
  
  The evolution of the atom-vacuum system is accompanied by two 
  simultaneous processes: spontaneous emission and associated atomic 
  recoil from the emitted photon.  The net result of these processes 
  is that, as noted before, the atomic and photonic states are 
  entangled \cite{guo10}:
  \begin{eqnarray}
  \label{e8}
  \vert \phi (t) \rangle &=&
  -\frac{1}{2\pi i} \int dq e^{-iqt/\hbar} 
  \int d\vec{p}_{0} \frac{f(\vec{p}_{0})}{q-p^{2}_{0}/(2m)-\hbar 
  \omega _{E} -B} \vert \vec{p}_{0} \rangle \otimes \vert E \rangle 
  \otimes \vert 0 \rangle \nonumber\\
  & & -\frac{1}{2\pi i} \int dq e^{-iqt/\hbar} \int 
  d\vec{p}_{0} \frac{f(\vec{p}_{0})}{q-p^{2}_{0}/(2m)-\hbar \omega _{E}
  -B} \nonumber\\
  & &\times \sum _{\alpha} \frac{ (\vec{\mu} _{GE} \cdot 
  \vec{g}_{\alpha})
  e^{-i\vec{k}_{\alpha}\cdot \vec{R}} \vert \vec{p}_{0} \rangle 
  \otimes \vert G \rangle 
  \otimes \vert 1_{\alpha} \rangle }{q-(\vec{p}_{0}-\hbar 
  \vec{k}_{\alpha })^{2}
  /(2m)-\hbar \omega _{G}-\hbar \omega _{\alpha}},
  \end{eqnarray}
  where $p_{0}=\vert \vec{p}_{0} \vert$, and 
  \begin{eqnarray}
  \label{9}
  B &\simeq &-\frac{\Gamma _{0}\hbar }{2\omega _{0} \pi}
  \Big [ \Omega +\omega _{0} \ln \big (\frac{\Omega -\omega _{0}}{
  \omega _{0}}\big )\Big ]-i \frac{\Gamma _{0}\hbar }{2}\nonumber\\
  &\equiv&B_{r}+iB_{i}.
  \end{eqnarray}
  In the expression of $B$, the quantity $\Gamma _{0}=4\mid 
  \vec{\mu}_{GE}\mid ^{2}\omega ^{3}_{0}/(3\hbar c^{3})$ is the 
  spontaneous emission rate of a stationary atom in the vacuum, and 
  $\Omega $ a cut-off frequency needed to make the nonrelativistic 
  Hamiltonian $H$ applicable in the present discussion 
  \cite{co89}.  Note, in Eq. (\ref{e8}), $(\vec{p}_{0}-\hbar 
  \vec{k}_{\alpha})^{2}/(2m)$ is the atom's kinetic energy after the 
  atom recoils from the photon emitted into mode $\vert 1_{\alpha}
  \rangle$.     
  
  Substitute Eq. (\ref{e8}) into Eq. (\ref{e7}) to get 
  \begin{equation} 
  \label{e10}
  \frac{d^{2}}{dt^{2}}\langle R_{x} \rangle (t) = -
  \frac{\omega ^{2}_{0}\vert \vec{\mu}_{GE} \vert ^{2}}{3m^{2}c^{5}}
  (\omega _{0}+B_{r}/\hbar )^{2} e^{-\Gamma _{0}t}
  \int d\vec{p}_{0} \vert f(\vec{p}_{0})\vert ^{2} p_{0}.
  \end{equation}
  In the derivation of Eq. (\ref{e10}), $\vec{\mu}_{GE}$ is averaged 
  over its orientation to conform to the fact that the orientation is 
  usually unknown.  Since the dependence of the atom's spontaneous 
  emission on the atom's speed is weak \cite{guo08}, the contribution 
  from the Doppler effect is ignored in the frequency of the emitted 
  photon $(\omega _{0}+B_{r}/\hbar)$.  Also used in Eq. (\ref{e10}) is 
  the mode-continuum approximation; see, for 
  example, Ref. \cite{guo05}.  A comparison of Eqs. (\ref{e7}) 
  and (\ref{e8}) shows that the derivative in Eq. (\ref{e10}) can never 
  survive unless the photonic and atomic states are entangled as in 
  Eq. (\ref{e8}).  
  
  Similarly, it is found that the second-order time derivatives of 
  $\langle R_{y} \rangle (t)$ and $\langle R_{z} \rangle (t)$ both 
  vanish, where $R_{y}$ and $R_{z}$ are the $y$- and $z$-components of 
  $\vec{R}$ respectively.  Thus, the radiation friction $\vec{F}_{r}$ 
  on the atom must be  
  \begin{eqnarray}
  \label{e11}
  \vec{F}_{r} &=& - \frac{\omega ^{2}_{0}\vert \vec{\mu}_{GE} 
  \vert ^{2}}{3mc^{5}} (\omega _{0}+B_{r}/\hbar )^{2} 
  e^{-\Gamma _{0}t} \int d\vec{p}_{0} \vert f(\vec{p}_{0})\vert
  ^{2} \vec{p}_{0} \nonumber\\
  &=&-\frac{(\omega _{0}+B_{r}/\hbar)^{2}\hbar \Gamma _{0}}{4
  m\omega _{0}c^{2}} 
  e^{-\Gamma _{0}t} \int d\vec{p}_{0} \vert f(\vec{p}_{0})\vert
  ^{2} \vec{p}_{0} .
  \end{eqnarray}
  
  Three properties of $\vec{F}_{r}$ are recognized.  First, the 
  magnitude of $\vec{F}_{r}$ decays with time exponentially.  This 
  observation is understandable, because the force is only present 
  when the atom and photon interact during spontaneous 
  emission, and spontaneous emission is largely an exponential 
  process.  For a discussion of spontaneous emission beyond the 
  rotating-wave approximation, see, for 
  example, Ref. \cite{guo05}.  Second, the force is proportional to 
  the average (initial) momentum of the atom, a character of the 
  average Langevin force if the atom's motion is viewed as a Brownian 
  motion \cite{pat96}.  Physically, the average momentum determines 
  the difference between the photonic frequencies when the photon is 
  emitted in or opposite to the direction of the atomic 
  velocity, and, thus, should also determine the strength of the 
  friction force.  A stationary atom, whose momentum is 
  zero, certainly does not experience the friction force 
  $\vec{F}_{r}$.  Finally, since it is proportional to the initial
  momentum of the atom, the force $\vec{F}_{r}$ must depend on the 
  coordinate system in which it is observed.  As Eq. (\ref{e7})
  shows, such dependence comes from the fact that $\frac{d^{2}}{d
  t^{2}}\langle R_{x} \rangle (t)$ is related to the wave vector
  $\vec{k}_{\alpha}$ of the photon, which, when combined with 
  $i\omega _{\alpha}/c$, is a 4-vector, and is different in different
  systems.  The friction force is not an invariant under the Lorentz
  transformation. 
  
  In conclusion, it is demonstrated that the motion of an excited atom 
  in the vacuum is subject to a friction force.  The force comes not
  only from the photon-atom entanglement but also from the momentum
  transferred from the emitted photon to the atom.

\end{document}